# MicroSim: Modeling the Swedish Population

Lisa Brouwers, Martin Camitz, Baki Cakici, Kalle Mäkilä, Paul Saretok

## Abstract

This article presents a unique, large-scale and spatially explicit microsimulation model that uses official anonymized register data collected from all individuals living in Sweden. Individuals are connected to households and workplaces and represent crucial links in the Swedish social contact network. This enables significant policy experiments in the domain of epidemic outbreaks. Development of the model started in 2004 at the Swedish Institute for Infectious Disease Control (SMI) in Solna, Sweden with the goal of creating a tool for testing the effects of intervention policies. These interventions include mass vaccination, targeted vaccination, isolation and social distancing. The model was initially designed for simulating smallpox outbreaks. In 2006, it was modified to support simulations of pandemic influenza. All nine millions members of the Swedish population are represented in the model. This article is a technical description of the simulation model; the input data, the simulation engine and the basic object types.

## Introduction

For policy purposes, understanding the behaviour of infectious diseases within a given population is vital. Traditionally SIR-models have been used to aid the understanding of these processes. SIR (Susceptible-Infected-Recovered) is a compartmental model in which differential equations govern the dynamic flow between three compartments and no contact structure is assumed. In an SIR-type model, the population is split into three different groups and the majority of the population is placed in the susceptible compartment. As the experiment progresses, individuals are randomly picked and moved between compartments according to the properties of the simulated disease. This lack of contact structure is referred to as random mixing. The individuals are considered to be exact copies of each other and no defining characteristics such as age, sex or social context are taken into consideration. All individuals are assumed to have the same number of contacts and are at equal risk of infection.

It has been demonstrated that simplified assumptions about contact structure can be sufficient for creating accurate transmission models (Andersson and May 1992, Diekmann and Heesterbeek 2000). This is especially true for highly infectious diseases such as measles. For diseases that are less infectious such as STDs or SARS, however, simplified compartmental models provide inadequate representations because contacts between susceptible and infectious persons are not random (Morris 1997, Liljeros et al. 2001, Dezső and Barabási 2002, Meyers 2007). Social structure, geography and behaviour are all important factors that influence the size and the speed of an outbreak. If individuals are represented explicitly in the model, the effect of contact structure can be observed and the change in individual behaviour due to policy or outbreak awareness can be simulated. Finally, by representing the geography explicitly, spatial concerns, constraints and their influence on the outbreak can be observed in silico.

## Input Data

All inhabitants of Sweden are assigned a 10-digit national registration number that serves as a unique personal identifier. The identifier is used in most administrative registries and provides an opportunity to unambiguously link information from different sources. The MicroSim model uses registry data obtained from Statistics Sweden (SCB) to generate the simulated population. The data is processed in two stages before it is parsed by MicroSim:

1. Data from three administrative registers is compiled into one base file by SCB.
2. The base file is transformed and modified into the format required for MicroSim. Three MicroSim-specific files (popfile, workplacefile, xyfile) are created.

### *Creating the Base File*

The data from three administrative registers is linked by the unique personal identifier. This linking and the substitution of an anonymized personal identifier instead of the real one is performed by SCB (Ethical approval identifier 04/903/2). The three registers used in this process are:
1. National population register
2. Employment register
3. Geography database

*National Population Register (2002)*

- **IndID**: Unique individual identifier.
- **BirthY**: Year of birth.
- **Sex**: 1 = Male, 2 = Female.
- **FamType**: 1 = Married/cohabiting with children, 2 = Married/cohabiting without children, 3 = Single with children, 4 = Single without children, 5 = Other.
- **FamID**: Family identifier.
- **WorkplaceID**: See the Employment Register section.
- **SchoolID**: Unique school identifier.
- **FatherID**: IndID of the individual's father.
- **MotherID**: IndID of the individual's mother.

*Employment Register (2002)*

- **CompanyID**: Unique company identifier.
- **WorkplaceID**: Unique workplace identifier. A workplace identifier uniquely identifies the workplace and the company.
- **WPbranchcode**: Branch code (1 to 1054).
- **WPmun**: The municipality of the workplace (1 to 290).

- **WPType**: 1 = Daycare center, 2 = School (ages 6 - 16), 3 = College (ages 16 - 19), 4 = Other school, 5 =Workplace.
- **SchoolID**: Unique school identifier.
- **SchoolType**: The highest level of education for the school (D = Daycare, L = Ages 6 - 9, M = Ages 10 - 12, H = Ages 13 – 15, HSK = College/University).

*Geographic database (2003)*

All objects in the population register and the employment register (households and workplaces) are aggregated into 100-meter squares. The eastern and northern coordinates are combined into one number where the first seven digits constitute the eastern coordinate and the following seven digits constitute the northern coordinate. For instance, the combined coordinates `15943006571200` represent `East 1594300` and `North 6571200`. These coordinates mark the lower left corner of the 100-meter square.

- **DwSquare**: Family household coordinates.
- **WPSquare**: Workplace coordinates.
- **SchoolSquare**: School coordinates.

*The compiled data file from SCB*

The base file from SCB contains one line per individual (see Table 1).

| IndId | BirthY | Sex | Fam Type | Dw Square | WP Square | School Square | FamID | Workpl. ID | WPbranch code | WP mun | Company ID | School ID | School Type | FatherID | MotherID |
|---|---|---|---|---|---|---|---|---|---|---|---|---|---|---|---|
| 2544587 | 1978 | 1 | 3 | 1594… | | | 2543583 | | | | | | | 2555897 | 2543583 |
| 2035042 | 1978 | 2 | 1 | 1623… | | 1335… | 2018599 | 999992 | | | 279737 | 022 | HSK | 2018599 | 2020201 |
| 1712109 | 1978 | 1 | 4 | 1621… | 1628… | | 1712109 | 200415 | 45230 | 1283 | 387818 | | | 1687754 | 1689503 |

*Table 1. Three SCB base file entries.*

### Creating the MicroSim files

The popfile is a binary file containing following variables: IndID, Sex, Age, FamID, DwellingSquare, WorkplaceID, BranchCode, and IndMunicipality. The workplace file contains the following information for each workplace: WorkplaceID, WPSquare, NoOfEmployees, Branch-code and Wpmun. In the internal representation the occupied grid squares are marked. The xy-file links these xy-indexes and the corresponding geographic coordinates.

## Data structure

### Objects

MicroSim uses two main object types to represent *persons* and *houses*. The house object is used to represents workplaces, schools, hospitals and homes. Each house or person object is associated with an executable *event* object called life_of_house or life_of_person and placed on the time line. The event and the corresponding person or house are implemented as separate objects to save space, since not all persons need to have an event object. The event object of a person takes care of day-to-day movement between home and workplace, to an emergency room (ER) or to a department of infectious disease (DID) if necessary. The event object of a house is used for disease transmission within the represented place by iterating through its member list.

All objects are stored in a fixed-size buffer in order to avoid the overhead of using the C++ *new* operator. The person and house objects are marked with a unique integer identifier which also contains information about the object type. The identifier is designed to quickly separate schools and kindergartens from hospitals, ERs and other care units.

### *Lists of members*

Every person is linked to a household and most are additionally linked to a workplace or a school. Information about the structure of this network is read from the three data files (pop, workplace, schoolkids) as described in the Data section. The house objects maintain arrays of their members. Each house member array contains up to four person objects which is sufficient for most homes and workplaces. When memory for additional members is required, chunks of 20 slots are allocated from a global fixed size person-buffer. These chunks can be allocated and released dynamically, for example, when modeling daily visits to ERs.

The person object stores the properties of the represented person's age and sex as well essential run time variables such as time of infection, the level of vaccination success and the current location.

The house object contains a standard code identifying the type of workplace, a region identifier and a geographic grid index. Certain flags enable the house to be removed from the timeline when there is no risk for its members to become infected. Extra data for care-type houses is contained in a separate structure that is associated with the house object using a hash-table.

### *Simulation engine*

The MicroSim simulation engine is event driven with discrete time steps along a timeline. The timeline array is composed of small event notice structures each representing a point in time. In the current implementation the time step is one hour. Each event notice in the timeline contains an index into the buffer where the actual executable event objects are stored.

Event notices scheduled for operation are stored in an array. A one-directional linked list is maintained for each timeline array slot, enabling the simulation engine to find the next scheduled event notice using a *next-index.* When a process is deactivated, it is not removed from the timeline but marked with a flag to prevent it from being executed. This avoids the overhead of removing the object and reordering the list. C-style pointers are not used in these sections of the source code in order to make allocation and garbage collection more efficient.

### *Geography*

For the purposes of our simulation, the entire resolution of the coordinate system is not required. By removing the large offset from origin and two trailing zeros (due to the 100-meter step size), the coordinate pair can be reduced to 32 bits. However, compared to the number of available 100-meter squares, only a very low number (~900000) are occupied, with most squares containing several homes and workplaces. A more efficient solution is numbering the grid squares and allowing the objects that require location information to store this xy-index. The link between the index and the actual coordinate is kept in an array which is read from a file when the simulation engine is initialized. Using this method, the xy-index can be stored in 20 bits and individuals can be assigned to care units (ERs and DIDs) more efficiently. An array linking the grid square index to the care units is used to determine the closest care unit.

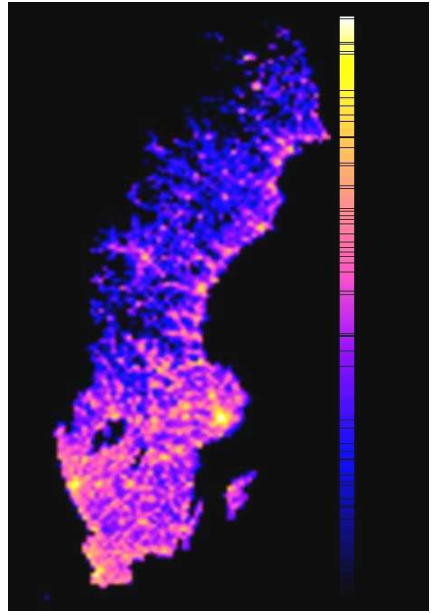

*Figure 1:* Population density map of Sweden.

### Storage/Memory

Microsimulation of 9 million individuals, their homes and workplaces is complex and memory intensive. Therefore, the simulation engine has been optimized to a high degree. This is achieved in most cases by bypassing the compiler's general purpose memory management and instead using a small set of customized memory management functions. In comparison, the built-in *new* function is slow and creates an extra overhead on the allocated objects, making them much larger than they need to be.

For each of the main object types, a sufficiently large buffer is allocated to store the objects. The objects are initialized by using the custom memory allocation function. Since they are accessed throughout the simulation, no garbage collection is required. The only module where the *new* operator is used is the alternative travel module which allocates several dynamic objects to a small portion of the population.

### Simulation

Simulation proceeds by iterating through the timeline array. At every step (corresponding to one hour), the list of events scheduled for that step are executed consecutively. After an event is run, its event notice is removed from the list to be reused for another event. All events for person objects are executed one step prior to those for house objects to ensure that persons are in the right position before disease transmission takes place.

*Disease Transmission*

The run method of a person object is in charge of day to day movement which is performed every day at 8 am and 4 pm. Depending on the time of day and current disease level of the person, there is a certain probability of the person being a) at home; b) at work, school or day care; c) at an ER in the vicinity; or d) at an infectious disease department of a nearby hospital. This is done by setting a pointer to the house representing the current location (transmission site). The pointer is checked when disease transmission takes place one hour later. Allocation to ERs and hospitals is done by calculating the Euclidean distance and searching for the closest ER or hospital. In most cases a clinic is already assigned to the particular grid square previously because the bulk of clinic-grid square association is read from a file at initialization.

Disease transmission is performed twice daily at 9 am and 5 pm. The house member lists and patient lists are iterated to calculate the combined infectiousness of their members. In the case of larger workplaces, the member lists are further divided into departments. After the combined infectiousness is determined, the list members are exposed and infected according to infection risk.

*Optimization*

The most significant performance increase is gained by keeping the number of active processes to a minimum. Only the infected individuals or those at risk of being infected need to be active on the timeline. The moment a person is infected, the house and workplace associated with that person are activated and placed on the timeline. Persons and houses are deactivated when they are no longer at risk i.e. immune or deceased. Houses with zero infection risk for a number of consecutive days are also deactivated. All objects are analysed every five days and deactivated when possible. These operations ensure that a minimum number of processes are active at any given moment during the simulations.

*Output*

During a simulation whenever an infection takes place a log entry is created. Each log entry contains simulation seeds, infection time, infection place type, infection place coordinates and infection risk. In addition, every entry also includes the id, home, region, age, sex, and the department for both the infector and the infected. The log files are stored as tab-separated plaintext files where each entry is terminated by a line break.

# Discussion

Explicit representation of social connections creates a population network that is suitable for realistic simulations of infectious disease outbreaks in Sweden. Most microsimulation models use either sample data or a fictive population, whereas MicroSim uses real register data. To our knowledge, MicroSim is the only large-scale population network model built on real register data. The model is not restricted to simulations of infectious disease, but could be used for all policy investigations where analysis of the distributional (social and/or geographical) effects is a priority.

# References

Andersson, R. M. & May, R. M. (1992). *Infectious Diseases of Humans: Dynamics and Control*, Oxford Univ Press, Oxford.

Dezső, Z. & Barabási, A.-L. (2002). "Halting viruses in scale-free networks", *Phys. Rev. E* **65**(5).

Diekmann, O. & Heesterbeek, J. (2002). *Mathematical Epidemiology of Infectious Diseases: Model building, analysis and interpretation*, John Wiley and Sons, Chichester.

Liljeros, F., Edling, C. R., Amaral, L. A. N. & Aberg, Y. (2001). "The web of human sexual contacts", *Nature* **411**, 907-908.

Meyers, L. A. (2007). "Contact network epidemiology: bond percolation applied to infectious disease prediction and control", *Bull. Amer. Math. Soc.* **44**, 63-86.

Morris, M. (1997). "Sexual networks and HIV", *AIDS* **11**, 209-216.

# Appendix: Object Attributes and Methods

**Person Object**
**Attributes**
actor_id, age, deceased, department, disease_profile, home, immune, life, sex, time_of_infection, transmission_site, workplace

**Methods**
choose_place, connect, decide_latent_period, disease_level, expose, host_state, infected, infectious, init, set_host_recovered

---

**House Object**
**Attributes**
actor_id, house_type, isInfected, life, members, region, sni_kod, xy

**Methods**
Add, capacity, compute_risk, get_occupancy, hospital_scan, increment_occupancy, infect_persons_callback_one, init, init_akutmott, init_home, init_inf_klinik, init_workplace, is_full, member_count, member_scan, print, transmit, wakeup, visit_akutmott, visit_inf_klinik , visitor_scan

---

**Event object (life_of_house/life_of_person)**
**Attributes**
CURRENT_PROCESS_ID, exec_state, houses_currently_active / persons_currently_active, MAX_PROCESS_ID, MAX_PROCESS_ID_WITHIN_BUFFER, myself, notice, process_id, see_events, self, to_be_passivated

**Methods**
Activate, allocate_array, cancel, delete_array, hold process, idle, init, maxProcessId, passivate process, passivateDelayed, passivating, run, set_size, terminate

---

**Time_axis object**
**Attributes**
cleanup_count, cleanup_limit, current, events, max_process_id, now, remove_count, running_limit

**Methods**
add_event,  add_event_after, add_event_before, add_event_now, find_next_event, Get_Time, print_all_events, remove_event_notice, remove_first, run, run_next_event, set_current, set_run_limit, set_size, time_axis